\documentclass[pra,aps,superscriptaddress,balancelastpage,onecolumn]{revtex4-2}
\usepackage{amsmath}
\usepackage{graphicx}
\usepackage{amsfonts}
\usepackage{amssymb}
\usepackage{bbold}

\providecommand{\U}[1]{\protect\rule{.1in}{.1in}}
\providecommand{\U}[1]{\protect\rule{.1in}{.1in}}

\begin{document}

\title{Enhanced bipartite entanglement and Gaussian quantum steering of squeezed magnon modes}

\author{Shaik Ahmed}
\affiliation{School of Technology, Woxsen University, Hyderabad, Telangana -502345, India}
\author{M. Amazioug}
\affiliation{LPTHE-Department of Physics, Faculty of sciences, Ibn Zohr University, Agadir, Morocco}
\author{Jia-Xin Peng}
\affiliation{State Key Laboratory of Precision Spectroscopy, Quantum Institute for Light and Atoms, Department of Physics, East China Normal University, Shanghai 200062, China}
\author{S. K. Singh\footnote{Corresponding Author}}
\email{singhshailendra3@gmail.com}
\affiliation{Graphene and Advanced 2D Materials Research Group, Sunway University, Malaysia}
\begin{abstract}
We theoretically investigate a scheme to entangle two squeezed magnon modes in a double cavity-magnon system, where both cavities are driven by a two-mode squeezed vacuum microwave field.  Each cavity contains an optical parametric amplifier as well as a macroscopic yttrium iron garnet (YIG) sphere placed near the maximum bias magnetic fields such that this leads to the excitation of the relevant magnon mode and its coupling with the corresponding cavity mode.  We have obtained optimal parameter regimes for achieving the strong magnon-magnon entanglement and also studied the effectiveness of this scheme towards the mismatch of both the cavity-magnon couplings and decay parameters.  We have also explored the entanglement transfer efficiency including Gaussian quantum steering in our proposed system.

\end{abstract}

\date{\today}

\maketitle

\section{Introduction}
Quantum entanglement \cite{Horod} and Gaussian quantum steering \cite{Rev, PRL115}  are two major important resources in the field of quantum computing \cite{comp}, quantum cryptography \cite{crypto} and quantum teleportation \cite{tele} including quantum information processing \cite{info}. Many microscopic as well as macroscopic quantum systems have been proposed over the past decades for the study of quantum entanglement and other nonclassical quantum correlations in superconducting qubits \cite{Neeley}, atomic ensembles \cite{juls}, cavity optomechanics \cite{Vittali,Peng,Sohail,Teklu,LID,Raymond,Sithi} and cavity magnomechanical (CMM) systems \cite{Hidki,Hidki1,Teklu1,Quent,Liao,JLi} which paves the way for advancements in present era of quantum technology. In CMM systems, the magnons defined as the collective excitation of a large number of spins in ferrimagnetic materials play very important role in the study of light-matter interactions  due to their  tunability, low damping, high spin density \cite{XZ,DZ} as well 
the strong coupling with the microwave photons \cite {Bai,Bourhill,Shahzad, YTAB}. Moreover,  other important mascroscopic  quantum  phenomena such as magnon-induced effects \cite{MNID}, tunable magnomechanically induced transparency and absorption \cite{Naseem,MIT}, slow light \cite{SL}, four-wave mixing \cite{FWM}, squeezed states \cite{Sque,Squee,Squeee,Squ}, nonclassical quantum correlations \cite{hidki2022quantifying,MPLA}, microwave-to-optical carrier conversion \cite{XZH} including quantum sensing \cite{Zhang, Lachance} also successfully investigated in 
cavity magnomechanical systems.\\
To  quantify the bipartite entanglement  between the magnon and 
microwave photon in CMM systems, we use  a very well-known witness of bipartite entanglement defined as the logarithmic negativity \cite{Adesso}. A recent theoretical work given in \cite{You}
explored the logarithmic negativity between two magnon modes where the optimal conditions for achieving the strong magnon-magnon entanglement involve the resonant coupling between the microwave cavity with both the magnon modes whereas in case of two microwave cavities given in \cite{JOSAA,AmjadShah,Hassan}, it is found that the detuning of the cavity and magnon mode   significantly  affects the bipartite entanglement. These research works also found the presence of both one-way and two-way  Gaussian quantum steering. So, all these studies  demonstrate that the bipartite entanglement and  quantum steering in CMM systems  can be significantly controlled through the various physical parameters. All these recent progress  also broaden our understanding of quantum correlations and facilitate to further explore the possibility of secure quantum protocols in such kind of macroscopic quantum systems.\\
Motivated by these works, we study the quantum correlations  and Gaussian quantum steering between two squeezed magnon modes of two yttrium iron garnet (YIG) spheres in a system of two spatially separated microwave  cavities.  Each cavity also contains an optical parametric amplifier (OPA) as well as a macroscopic YIG sphere placed near the maximum bias magnetic fields such that this leads to the excitation of the corresponding magnon mode and its coupling with the cavity mode. In addition, both the cavities are simultaneously driven by a two mode squeezed vacuum microwave field in our proposed  system \cite{Tabu}. In this present  work, we found the generation of a considerable bipartite entanglement between the two magnon modes  with gradually increasing squeezing parameter and the mean thermal magnon number. Moreover, it can be seen clearly from our work that not all entangled states allow for quantum steering whereas any state that can be steered must necessarily be entangled.\\
This paper is organized as follows: In Section II, we introduce model Hamiltonian and also evaluate corresponding  quantum Langevin equations including its solutions (QLEs). In Section III, we discuss in details about mathematical formulation of bipartite entanglement and Gaussian quantum steering between two magnon modes. Numerical Results and related  discussions are given in Section IV whereas we conclude our results in Section V.

\section{The Model Hamiltonian}
Our proposed system shown in Fig. 1 consists of two microwave cavities and two magnon modes in two YIG spheres,
which are respectively placed inside the cavities near the maximum magnetic fields of the cavity modes and simultaneously in
uniform bias magnetic fields such that it causes both the magnon modes to strongly couple with respective  cavity modes \cite{Strong1, Tabu}. Each cavity, contains a degenerate optical parametric amplifier (OPA) to produce squeezed light \cite{DFWalls1994}. The Hamiltonian of the system in a rotating frame with frequency $\omega_{j}$ can be written as

\begin{equation} \label{Hamilt}
\mathcal{H} =\underset{j=1,2}{\sum }\Big\{\hbar\Delta _{c_{j}}c_{j}^{\dagger}c_{j}+\hbar\Delta _{m_{j}}m_{j}^{\dagger}m_{j}+\hbar g_{j} \big(c_{j}m_{j}^{\dagger}+c_{j}^{\dagger}m_{j}\big)+ic \hbar \lambda_{j}(e^{ic\theta}c_{j}^{\dagger 2}-e^{-ic\theta}c_{j}^{2}) + ic\hbar \mu_{j} (e^{ic \nu}m_{j}^{\dagger 2}-e^{-ic \nu}m_{j}^{2})\Big\},
\end{equation}
where $\Delta_{c_{j}}=\omega_{c_{j}}-\omega_{_{j}},\Delta_{m_{j}}=\omega _{m_{j}}-\omega_{_{j}}$, $c_{j}$ ($c_{j}^{\dagger}$), $m_{j}$ ($m_{j}^{\dagger}$) are the annihilation (creation) operators of the $j^{th}$ cavity and magnon modes, respectively, and we have $\big[\mathcal{O}, \mathcal{O}^+\big] =1$ ($\mathcal{O}\,{=}\,c_{j}, m_{j}$).  $\omega _{c_{j}}$ ($\omega_{m_{j}}$) is the resonance frequency of the $j^{th}$ cavity mode (magnon mode). The frequency of the magnon mode $\omega _{m_{j}}$ is determined by the external bias magnetic field $H_{j}$ and the gyromagnetic ratio $\beta$ via $\omega _{m_{j}}=\beta H_{j}$, and thus can be flexibly adjusted, and  $g_{j}$ is the coupling rate between the $j^{th}$\ cavity and magnon modes. The parameter $\lambda_j$ and $\theta_j$ represents respectively the nonlinear gain of the OPA and the phase of the driving field.  with $\mu_j$ being the squeezing parameter and $\nu$ being the phase of $j^{th}$ squeezing mode. The magnon squeezing can be achieved by transferring squeezing from a squeezed-vacuum microwave field \cite{FWM}, or by the intrinsic nonlinearity of the magnetostriction (the so-called ponderomotive-like squeezing) \cite{JLiarxiv}, or by the anisotropy of the ferromagnet \cite{HYYuanarxiv,AKamara2016}, etc.  

\begin{figure}[htb]\label{fig1}
\hskip-1.0cm\includegraphics[width=0.95\linewidth]{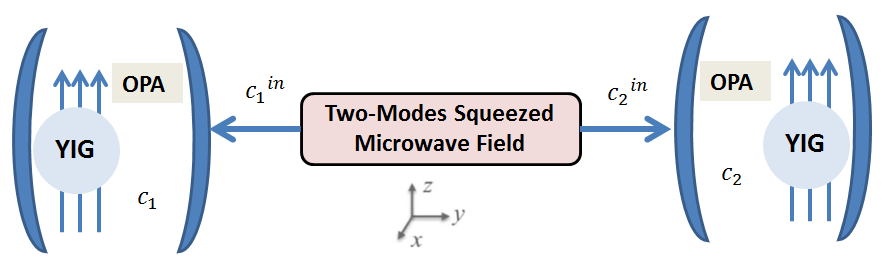} 
\caption{Schematic diagram of a double cavity-magnon system where both the cavities are driven by a two-mode squeezed vacuum microwave field. Two YIG spheres are respectively placed inside the microcavities near the maximum magnetic fields of the cavity modes and simultaneously in uniform bias magnetic fields. }
\end{figure}

In the frame rotating at the frequency $\omega _{_{j}}$, i.e., the frequency of the $j^{th}$ mode of the input two-mode squeezed field, the QLEs of this model Hamiltonian are given by
\begin{eqnarray} \label{QLEs}
\dot{c}_{j} &=&-(\kappa_{c_{j}}+i\Delta_{c_{j}})a_{j}-ig_{j}m_{j}+2\lambda_{j}e^{i\theta} c^{\dagger}{}_j+\sqrt{2\kappa _{c_{j}}}c_{j}^{in},  \\
\dot{m}_{j} &=&-(\kappa_{m_{j}}+i\Delta_{m_{j}})m_{j}-ig_{j}c_{j}+2\mu_j e^{ic \nu} m^{\dagger}_{j}+\sqrt{2\kappa_{m_{j}}}m_{j}^{in},  \nonumber
\end{eqnarray}
where $\kappa _{c_{j}}$ ($\kappa _{m_{j}}$) is the decay rate of the $j$th cavity mode (magnon mode), $\Delta _{c_{j}}=\omega_{c_{j}}-\omega _{_{j}},\Delta _{m_{j}}=\omega _{m_{j}}-\omega _{_{j}}$, and $c_{j}^{in}$ ($m_{j}^{in}$) is the input noise operator for the $j^{th}$ cavity mode (magnon mode). The two cavity input noise operators $c_j^{in}$ are quantum correlated due to the injection of the two-mode squeezed field, and have the following non-zero correlations in time domain $c^{in}_{j}$ and $c^{in\dagger}_{j}$ are given by \cite{JLi2020} 
\begin{equation} \label{eq:6} 
\langle c^{in}_{j}(t)c^{in\dagger}_{j}(t')\rangle = (\mathcal{N}+1)\delta(t-t')
\end{equation}
\begin{equation} \label{eq:7} 
\langle c^{in\dagger}_{j}(t)c^{in}_{j}(t')\rangle = \mathcal{N}\delta(t-t')
\end{equation}
\begin{equation} \label{eq:8} 
\langle c^{in}_{j}(t)c^{in}_{j'}(t')\rangle =  \mathcal{M}e^{-ic\omega_{M}(t+t')}\delta(t-t')\quad;\quad j\neq j'
\end{equation}
\begin{equation} \label{eq:9} 
\langle c^{in\dagger}_{j}(t)c^{in\dagger}_{j'}(t')\rangle =  \mathcal{M}e^{ic\omega_{M}( t+t')}\delta(t-t')\quad;\quad j\neq j'
\end{equation}.
Here $\mathcal{N}=\sinh^{2}r$, $\mathcal{M}=\sinh r\cosh r$ and $r$ is the squeezing parameter of the two-mode squeezed vacuum field whereas the magnon input noise operators $m_j^{in}$ are of zero mean and correlated as follows
\begin{equation} \label{eq:4} 
\langle m_{j}^{in}(t) m_{j}^{in\dagger}(t') \rangle = (N_{m_{j}}+1)\delta(t-t')
\end{equation}
\begin{equation} \label{eq:5} 
\langle m_{j}^{in\dagger}(t)m_{j}^{in}(t')\rangle = N_{m_{j}}\delta(t-t')
\end{equation}
where $N_{m_{j}}=\left[\exp \left(\frac{\hbar \omega _{m_{j}}}{k_{B}T}\right)-1\right]^{-1}$ is the equilibrium mean thermal magnon number of the $j^{th}$ mode, with $T$ the environmental temperature and $k_B$ the Boltzmann constant.

Using the linearisation of quantum Langevin equations, the fluctuations of the system are written as 

\begin{eqnarray}
\delta \dot{c}_{j} &=&-(\kappa _{c_{j}}+i\Delta _{c_{j}})\delta c_{j}-ig_{j}\delta m_{j}+2\lambda_j e^{i\theta}\delta{c}^{\dagger}_{j}+\sqrt{2\kappa _{c_{j}}}c_{j}^{in},  \label{QLEs2} \\
\delta \dot{m}_{j} &=&-(\kappa _{m_{j}}+i\Delta _{m_{j}})\delta m_{j}-ig_{j}\delta c_{j}+2\mu_{j}e^{ic \nu}\delta{m}^{\dagger}_{j}+\sqrt{2\kappa_{m_{j}}} m_{j}^{in}.  \notag
\end{eqnarray}
To get the explicit expression of the degree of freedom of optical and magnon modes, we consider the EPR-type quadrature fluctuations operators corresponding to the two subsystems defined as $\delta Q_{j}=(\delta c_{j}+\delta c_{j}^{\dag })/\sqrt{2},\delta P_{j}=i(\delta c_{j}^{\dag }-\delta c_{j})/\sqrt{2},\delta q_{j}=(\delta m_{j}+\delta m_{j}^{\dag })/\sqrt{2},\delta p_{j}=i(\delta m_{j}^{\dag }-\delta m_{j})/\sqrt{2}$ (we have similar definition for input noises $Q_{j}^{in}, P_{j}^{in}$ and $q_{j}^{in}, p_{j}^{in}$) \cite{EPJPlus,MDPI,Front,QIP,QIP1,PLA}, 

The above QLEs can be simplified as
\begin{eqnarray}
\delta \dot{Q}_{j} &=&-\kappa _{c_{j}}\delta Q_{j}+\Delta
_{c_{j}}\delta P_{j}+g_{j}\delta p_{j}+2\lambda\cos(\theta)\delta q_{j}+2\lambda\sin(\theta)\delta p_{j}+\sqrt{2\kappa _{c_{j}}}Q_{j}^{in},
\label{QELs3} \\
\delta \dot{P}_{j} &=&-\kappa _{c_{j}}\delta P_{j}-\Delta
_{c_{j}}\delta Q_{j}-g_{j}\delta q_{j}-2\lambda\cos(\theta)\delta p_{j}+2\lambda\sin(\theta)\delta q_{j}+\sqrt{2\kappa _{c_{j}}}P_{j}^{in}, 
\notag \\
\delta \dot{q}_{j} &=&-\kappa _{m_{j}}\delta q_{j}+\Delta
_{m_{j}}\delta p_{j}+g_{j}\delta P_{j}+2\mu\cos(\nu)\delta Q_{j}+2\mu\sin(\nu)\delta P_{j}+\sqrt{2\kappa _{m_{j}}}q_{j}^{in}, 
\notag \\
\delta \dot{p}_{j} &=&-\kappa _{m_{j}}\delta p_{j}-\Delta
_{m_{j}}\delta p_{j}-g_{j}\delta Q_{j}-2\mu\cos(\nu)\delta P_{j}+2\mu\sin(\nu)\delta Q_{j}+\sqrt{2\kappa _{m_{j}}}p_{j}^{in}. 
\notag
\end{eqnarray}

Equation (\ref{QELs3}) take the following compact matrix form 
\begin{equation}
\dot{V}(t)=\mathcal{A} V(t)+\chi (t),  \label{MatrixForm}
\end{equation}
Here $V(t)=[\delta Q_{1},\delta P_{1},\delta Q_{2},\delta P_{2},\delta q_{1,}\delta p_{1},\delta q_{2},\delta p_{2}]^{T}$, $\mathcal{A}$\ is the drift matrix

\begin{equation} \label{driftmatrix:A}
\mathcal{A}=
\begin{pmatrix}
	    \mathcal{A}_1 & \mathcal{A}_3  \\
    \mathcal{A}_3 & \mathcal{A}_2 
\end{pmatrix}
\end{equation} 

where

\begin{equation} \label{eq:A1}
\mathcal{A}_1=
\begin{pmatrix}
	    -\kappa _{c_{1}}+2\lambda\cos(\theta) & \Delta _{c_{1}}+2\lambda\sin(\theta) & 0 & 0  \\
    -\Delta _{c_{1}}+2\lambda\sin(\theta) & -\kappa _{c_{1}}-2\lambda\cos(\theta) & 0 & 0   \\
    0 & 0 & -\kappa _{c_{2}}+2\lambda\cos(\theta) & \Delta _{c_{2}}+2\lambda\sin(\theta)   \\
    0 & 0 & -\Delta _{c_{2}}+2\lambda\sin(\theta) & -\kappa _{c_{2}}-2\lambda\cos(\theta) 
\end{pmatrix}
\end{equation} 

and

\begin{equation} \label{eq:A2}
\mathcal{A}_2 =
\begin{pmatrix}
	    -\kappa _{m_{1}}+2\mu\cos(\nu) & \Delta _{m_{1}}+2\mu\sin(\nu) & 0 & 0  \\
    -\Delta _{m_{1}}+2\mu\sin(\nu) & -\kappa _{m_{1}}-2\mu\cos(\nu) & 0 & 0   \\
    0 & 0 & -\kappa _{m_{2}}+2\mu\cos(\nu) & \Delta _{m_{2}}+2\mu\sin(\nu)   \\
    0 & 0 & -\Delta _{m_{2}}+2\mu\sin(\nu) & -\kappa _{m_{2}}-2\mu\cos(\nu) 
\end{pmatrix}
\end{equation} 

and

\begin{equation} \label{eq:A3}
\mathcal{A}_3=
\begin{pmatrix}
	0 & g_1 & 0  & 0 \\
    -g_1 & 0   & 0  & 0 \\
    0 & 0 & 0  & g_2 \\
    0 & 0   & -g_2  & 0
\end{pmatrix}
\end{equation} 
and $\chi(t)=[\sqrt{2\kappa _{c_{1}}}Q_{1}^{in},\sqrt{2\kappa _{c_{1}}}P_{1}^{in}, \sqrt{2\kappa _{c_{2}}}Q_{2}^{in},\sqrt{2\kappa _{c_{2}}}P_{2}^{in},\sqrt{2\kappa _{m_{1}}}q_{1}^{in},\sqrt{2\kappa _{m_{1}}}p_{1}^{in},\sqrt{2\kappa_{m_{2}}}q_{2}^{in},\sqrt{2\kappa _{m_{2}}}p_{2}^{in}]^{T}$. 
The system is stable when eigenvalues of the drift matrix $\mathcal{A}$ (\ref{driftmatrix:A}) have negative real parts. This corresponds to the so-called Routh-Hurwitz criterion \cite{EXDeJesus1987}. The steady state of the system, which is completely characterized by an $8\times 8$ covariance matrix (CM) $\Sigma$, defined as $\Sigma_{ij}(t)=\langle V_{i}(t)V_{j}(t^{\prime })+V_{j}(t^{\prime })V_{i}(t) \rangle/2$. the solution of $\Sigma$ can be obtained by directly solving the Lyapunov equation~\cite{DVitali2007} 
\begin{equation}
\mathcal{A} \Sigma+\Sigma\mathcal{A}^T=-\cal{D},  \label{LyapEq}
\end{equation}
where $\mathcal{D}$ is the diffuse matrix defined by $\mathcal{D}_{ij}\delta (t-t^{\prime})=\langle \chi_{i}(t) \chi_{j}(t^{\prime })+\chi_{j}(t^{\prime })\chi_{i}(t) \rangle/2$, given by
\begin{equation} \label{eq:D}
\mathcal{D}=
\begin{pmatrix}
	    \kappa' & 0 & \sqrt{\kappa_{c_1}\kappa_{c_2}}\mathcal{M} & 0 & 0 & 0 & 0 & 0 \\
    0 & \kappa' & 0 & -\sqrt{\kappa_{c_1}\kappa_{c_2}}\mathcal{M} & 0 & 0 & 0 & 0   \\
    \sqrt{\kappa_{c_1}\kappa_{c_2}}\mathcal{M} & 0 & \kappa'' & 0 & 0 & 0 & 0 & 0  \\
    0 & -\sqrt{\kappa_{c_1}\kappa_{c_2}}\mathcal{M} & 0 & \kappa'' & 0 & 0 & 0 & 0 \\ 
		 0 & 0 & 0 & 0 & \gamma' & 0 & 0 & 0 \\
    0 & 0 & 0 & 0 & 0 & \gamma' & 0 & 0 \\
    0 & 0 & 0 & 0 & 0 & 0 & \gamma'' & 0  \\
    0 & 0 & 0 & 0 & 0 & 0 & 0 & \gamma''
\end{pmatrix}
\end{equation} 
where $\kappa'=\kappa_{c_1}\big(\mathcal{N}+\frac{1}{2}\big)$, $\kappa''=\kappa_{c_2}\big(\mathcal{N}+\frac{1}{2}\big)$, $\gamma'=\kappa _{m_1}\big(2N_{m_{1}}+1\big)$ and $\gamma''=\kappa _{m_2}\big(2N_{m_{2}}+1\big)$.

The covariance matrix $\Sigma_{(mm)}$ associated with the two magnon modes is given by

\begin{equation} \label{eq:Sigmamm}
\Sigma_{(mm)}=
\begin{pmatrix}
	\mathcal{X} & \mathcal{Z}  \\
    \mathcal{Z}^T & \mathcal{Y}  
\end{pmatrix}
\end{equation} 
The $2\times 2$ sub-matrices $\mathcal{X}$ and $\mathcal{Y}$ in Eq. (\ref{eq:Sigmamm}) describe the autocorrelations of the two magnon modes and $2\times 2$ sub-matrix $\mathcal{Z}$ in Eq. (\ref{eq:Sigmamm}) denotes the cross-correlations of the two magnon modes.

\section{Quantum correlations}

\subsection{Quantum entanglement}
The logarithmic negativity $E_m$ is a measure or witness of entanglement in bipartite continuous-variable (CV) systems \cite{GAdesso2004}. Mathematically, it can be expressed as:
\begin{equation} \label{eq:37}
	E_{m}=\max[0,-\log(2\psi^-)]
\end{equation}
with $\psi^-$ being the smallest symplectic eigenvalue of partial transposed covariance matrix $\Sigma_{(mm)}$ of two magnon modes 
\begin{equation} \label{eq:38}
\psi^-= \sqrt{\frac{\Gamma-\sqrt{\Gamma^2-4\det\Sigma_{(mm)}}}{2}}   
\end{equation}
where the symbol $\Gamma$ is written as $\Gamma=\det \mathcal{X}+\det \mathcal{Y}-\det \mathcal{Z}$. The two magnon modes are entangled if the condition $\psi^-<1/2$ (i.e. when $E_{m}>0$) is satisfied.

\subsection{Gaussian quantum steering}
Quantum steering is the process of acquiring information about an
unmeasurable quantum system by measuring a single quantum system.  Gaussian quantum steering is the asymmetric property between two entangled observers (the two mechanical modes), Alice ($A$: magnon $M_1$) and Bob ($B$: magnon $M_2$). Besides, they are used to quantify how much the two magnon modes are steerable. We use the covariance matrix $\Sigma_{(mm)}$ of the two magnon modes, the Gaussian steering $A\to B$ and $B\to A$ written as \cite{Ioan,Sohail} 
\begin{equation} \label{eq:36}
S^{A \to B}=S^{B \to A}=\max\left[0,\frac{1}{2}\ln{\left(\frac{\det(\mathcal{X})}{4\det\Sigma_{(mm)}}\right)}\right];
\end{equation}
There are two possibilities of steerability between $A$ and $B$ : (i) no-way steering if $S^{A \to B}=S^{B \to A}=0$ i.e. Alice can't steer Bob and vice versa also impossible even if they are not separable, and (ii) two-way steering if $S^{A \to B}=S^{B \to A}>0$, i.e. Alice can steer Bob and vice versa. Indeed, a non-separable state is not always a steerable state, while a steerable state is always not separable.

\section{Results and Discussion}
In this section, we will discuss the steady state quantum correlations of two magnon modes under various parameters regime reported experimentally \cite{Tabu,JLi2020} and given as $\omega_{c_1}/2\pi= 10$ GHz, $\kappa_{c}/2\pi= 5$ MHz, $\kappa_{m}=\kappa_{c}/5$, $g_1 = g_2 = 5\kappa_{c}$, $\theta=\pi$ and $\nu=0.9\pi$. For simplicity, we consider that $N_{m_1}=N_{m_2}=N_m$, $\kappa_{c_1}=\kappa_{c_2}=\kappa_{c}$ and $\kappa_{m_1}=\kappa_{m_2}=\kappa_{m}$. Additionally, each YIG sphere used in our study has a diameter of 0.5 mm. These spheres are specifically chosen for their size and contain more than $10^{17}$ spins. 
\begin{figure}[!htb]
\minipage{0.5\textwidth}
  \includegraphics[width=\linewidth]{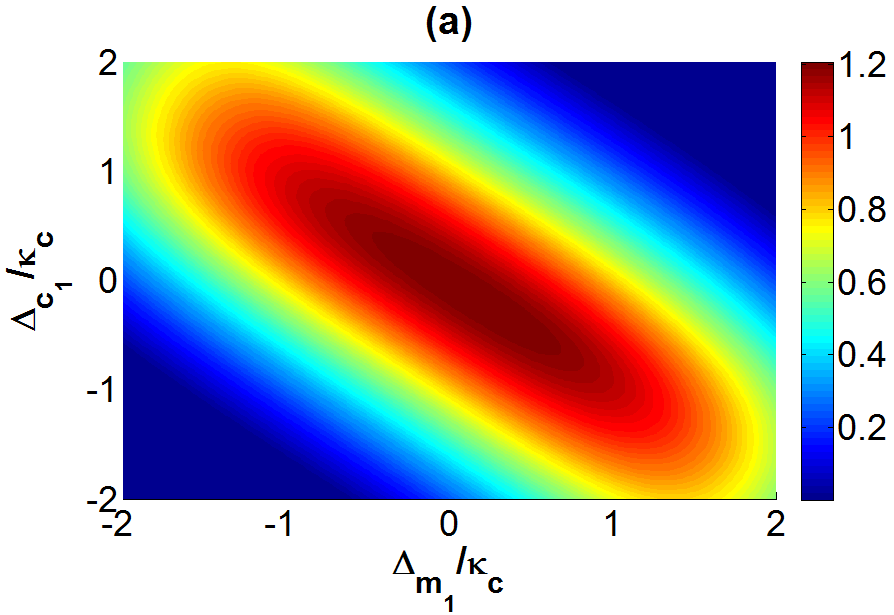}
%  \caption{A really Awesome Image}\label{fig:awesome_image1}
\endminipage\hfill
\minipage{0.5\textwidth}
  \includegraphics[width=\linewidth]{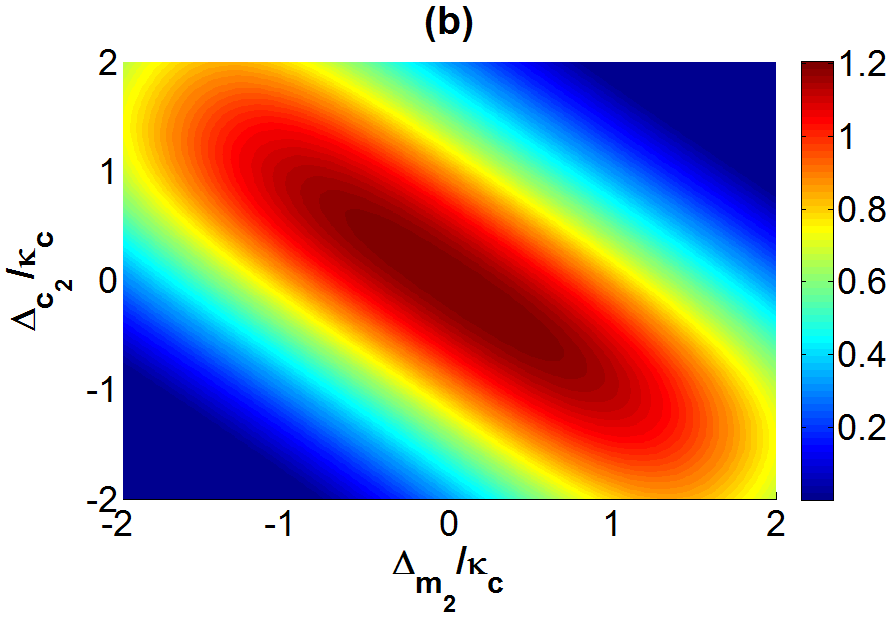}
 % \caption{A really Awesome Image}\label{fig:awesome_image2}
\endminipage\hfill
\caption{Plot of the logarithmic negativity $E_m$ between two magnon modes versus (a) $\Delta_{c_1}$ and $\Delta_{m_1}$, with $\Delta_{c_2}=\Delta_{m_2}=0$; (b) $\Delta_{c_2}$ and $\Delta_{m_2}$, with $\Delta_{c_1}=\Delta_{m_1}=0$, $r = 1$, $\lambda=0.2 \kappa_c$, $\mu=0.2 \kappa_c$ and $T = 100$ mK. See text for the other parameters.}
\label{CPB1}
\end{figure}

In Fig. \ref{CPB1}, we have plotted the logarithmic negativity $E_m$ of subsystem magnon-magnon with varying  $\Delta_{c_{j}}$ and $\Delta_{m_{j}}$ ($j=1, 2$) whereas all other parameters are fixed. It can be seen  in Figs. \ref{CPB1}(a) and (b) that when $\Delta_{c_{j}}=\Delta_{m_{j}}=0$ ($j=1, 2$), the entanglement between two magnon modes is optimal.  This observation can be attributed to the resonant transfer of quantum correlations from the input fields to the two magnon modes, facilitated by the linear cavity-magnon coupling. 

\begin{figure}[!htb]
\minipage{0.5\textwidth}
  \includegraphics[width=\linewidth]{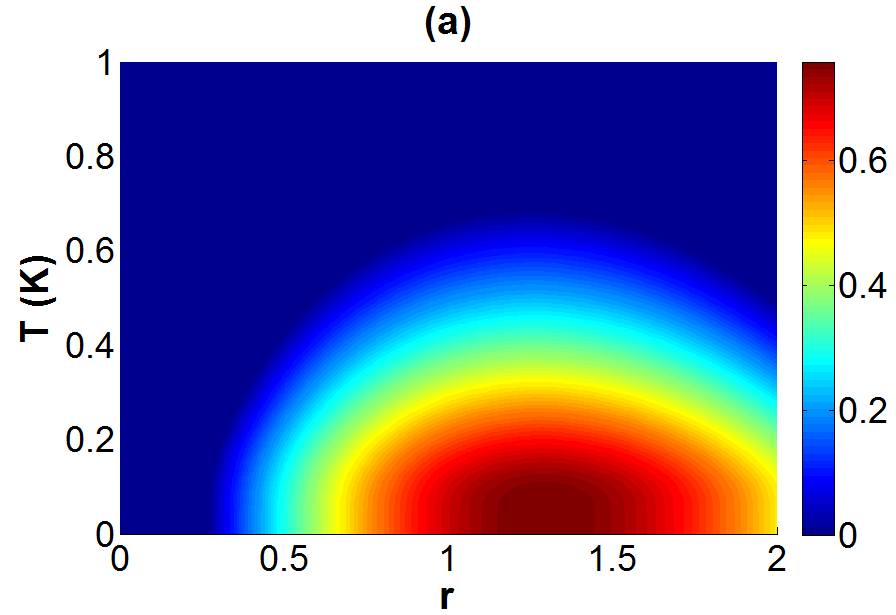}
%  \caption{A really Awesome Image}\label{fig:awesome_image1}
\endminipage\hfill
\minipage{0.5\textwidth}
\includegraphics[width=\linewidth]{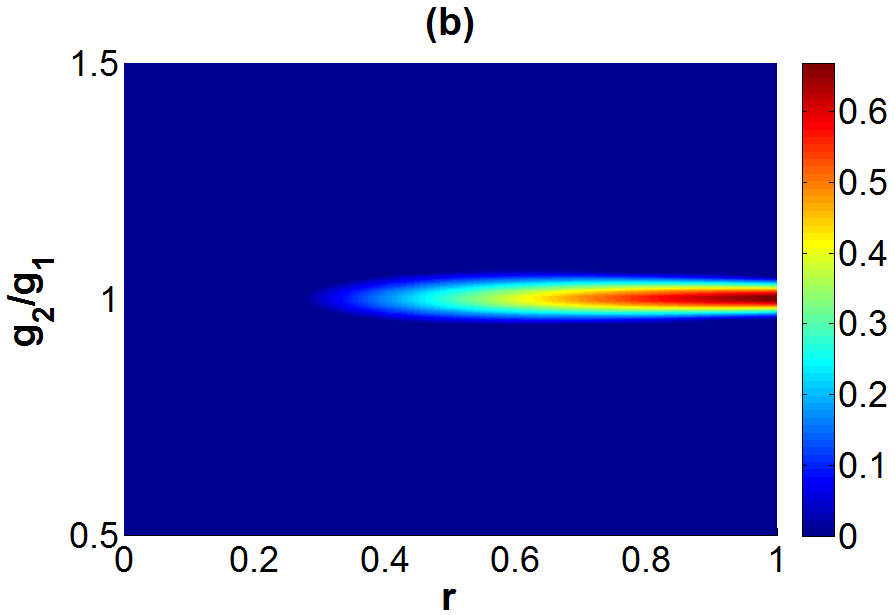}
\endminipage\hfill
\caption{(a) Plot of the logarithmic negativity $Emm$ between the two magnon modes vs  the temperature $T(K)$ and the magnon squeezing parameter $\mu$ with $\Delta_{c_{1,2}}=\Delta_{m_{1,2}}=0$, $T = 100$ mK, $\lambda=0.2\kappa_c$ and $\mu=0.2\kappa_c$. (b) Plot of the logarithmic negativity $Emm$ between the two magnon modes vs $g_2/g_1$ and $r$ of the two-mode squeezed vacuum field with $g_1 = 5\kappa_c$, $\Delta_{c_{1,2}}=\Delta_{m_{1,2}}=0$, $T = 100$ mK, $\lambda=0.2 \kappa_c$ and $\mu=0.2 \kappa_c$.}
\label{CPB2}
\end{figure}

Fig. \ref{CPB2}(a) shows that the bipartite entanglement $E_{m}$  increases with the squeezing parameter $r$ of the two-mode input squeezed vacuum field and decreases with the temperature $T$. The effect of the temperature is due to the influence of the decoherence phenomenon \cite{Decoherence2003}. Furthermore, it has been observed that the logarithmic negativity $E_{m}$ reaches its maximum value when the value of $r$ falls within the range of (1-1.5)  whereas when the parameter $r$ goes to zero both the magnon modes remain separable ($E_{m}=0$) as illustrated in figure \ref{CPB2}(a). This shows the dependence of the bipartite entanglement of the two magnon modes on the squeezing parameter $r$. In Fig. \ref{CPB2}(b), we plot $E_{m}$ as a function of squeezing parameter $r$ and $g_{2}/g_{1}$. We found here the generation of the bipartite entanglement between the two magnon modes with a gradual increase in squeezing parameter $r$ even for a wide range of mismatch of the two coupling strengths  $g_{2}$ and $g_{1}$.

\begin{figure}[!htb]
\minipage{0.315\textwidth}
  \includegraphics[width=\linewidth]{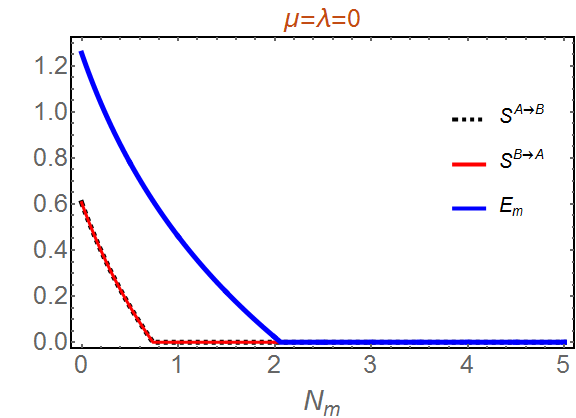}
%  \caption{A really Awesome Image}\label{fig:awesome_image1}
\endminipage\hfill
\minipage{0.315\textwidth}
  \includegraphics[width=\linewidth]{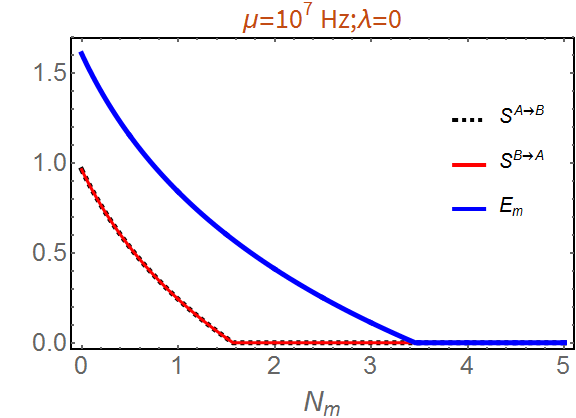}
 % \caption{A really Awesome Image}\label{fig:awesome_image2}
\endminipage\hfill
\minipage{0.315\textwidth}%
  \includegraphics[width=\linewidth]{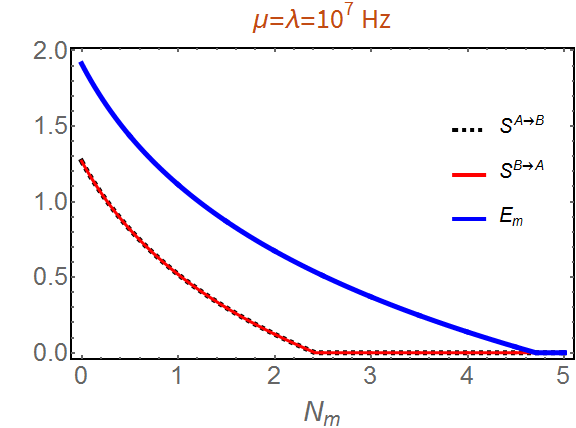}
 % \caption{A really Awesome Image}\label{fig:awesome_image3}
\endminipage
\caption{Plots of the steering $S^{A\to B}$, $S^{B\to A}$ logarithmic negativity $E_m$ of the two magnon modes versus the equilibrium mean thermal magnon number $N_m$ for various values of the coupling $\lambda$ and $\mu$, with $\Delta_{c_{1,2}}=\Delta_{m_{1,2}}=0$, $r = 1$ and $T = 100$ mK.}
\label{CPB3}
\end{figure}

In Fig.(\ref{CPB3}), we plot the Gaussian steering $S^{A\to B}$, $S^{B\to A}$ and logarithmic negativity $E_{m}$ for the subsystem magnon-magnon as a function of the equilibrium mean thermal magnon number $N_{m}$ for various values of the parameters $\lambda$ and $\mu$ whereas the other parameters are fixed.  It  can be
seen that $S^{A\to B}$, $S^{B\to A}$ and entanglement $E_{m}$ have the same evolution behavior. This figure studies the effect of $N_{m}$ (temperature $T$) and the parameters $\lambda$ and $\mu$ on the bipartite entanglement and quantum steerings.  Due to decoherence phenomena both the quantities i.e. bipartite entanglement and quantum steering decrease very quickly with increasing $N_{m}$. Moreover, when we enhance $\lambda$ and $\mu$ the magnon-magnon entanglement as well as two-way quantum steering become finite for a wide range of temperature $T$ (the equilibrium mean thermal magnon number $N_m$ ). Moreover, as depicted in Fig.(\ref{CPB3}) the entangled state is not always a steerable state but a steerable state must be entangled, i.e. when $S^{A\to B} = S^{B\to A} > 0$ which leads to $E_{m} > 0$ and hence is the witnesses of the existence of Gaussian two-way steering. This means that  both magnon modes are entangled  as well as are steerable  from $A$ to $B$ and from $B$ to $A$. However, we get no-way steering when $S^{A\to B} = S^{B\to A} = 0$ and $E_{m} > 0$ and so  the measure of Gaussian steering always remains bounded by the bipartite entanglement $E_{m}$.
\begin{figure}[!htb]
\minipage{0.5\textwidth}
  \includegraphics[width=\linewidth]{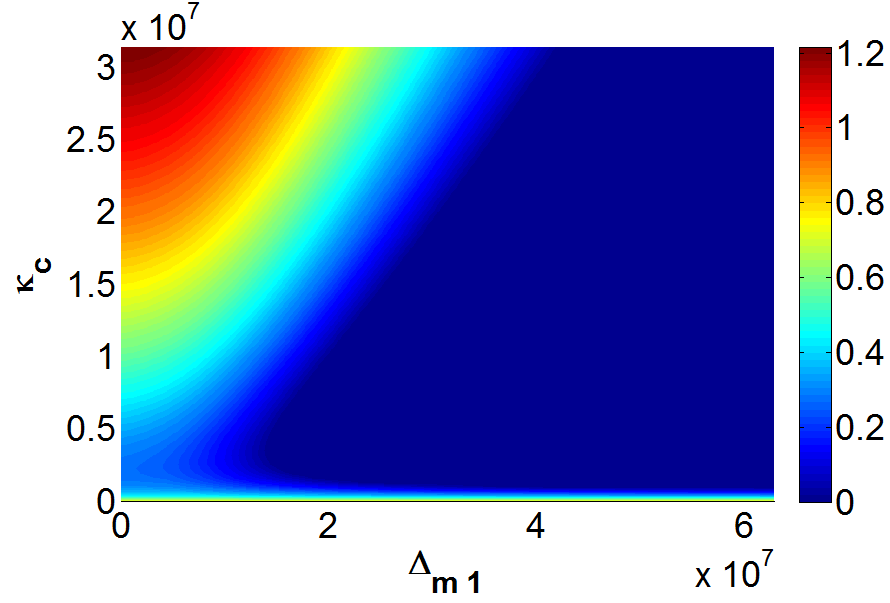}
%  \caption{A really Awesome Image}\label{fig:awesome_image1}
\endminipage\hfill
 
\caption{Plot of the logarithmic negativity $E_{m}$ between the two magnon modes vs $\kappa_c$ and $\Delta_{m_1}$ with $\Delta_{c_{1}}=\Delta_{c_{2}}=\Delta_{m_{2}}=0$, $T = 50$ mK, $r=1.5$, $g_{1} = 5\kappa_{c}$, $\lambda=0.2\kappa_{c}$ and $\mu=0.2\kappa_c$.}
\label{CPB4}
\end{figure}\\
In Fig. \ref{CPB4}, we have plotted the logarithmic negativity $E_{m}$ of two magnon modes with  $\kappa_{c}$ and $\Delta_{m_{1}}$ for a fixed value of all other parameters. It can be seen that the entanglement between the two magnon modes is maximum when $\Delta_{m_{1}}=0$ and $\kappa_{c}=3\times 10^7$ Hz. However, the bipartite entanglement $E_{m}$ decreases with decreasing decay rate $\kappa_{c}$ and increasing detuning $\Delta_{m_{1}}$.

\section{Conclusions}
We have theoretically investigated a scheme for the generation of the bipartite entanglement and Gaussian quantum steering in a double microwave cavity-magnon hybrid system where a two-mode squeezed microwave vacuum field is also transferred simultaneously into both cavities. We have obtained optimal parameter regimes for achieving the strong magnon-magnon entanglement and also explored the effectiveness of the scheme towards the mismatch of two cavity magnon couplings including the entanglement transfer efficiency. Our present study of bipartite entangled states of two magnon modes in coupled microwave resonators has important applications in coherent control of various non classical correlations in macroscopic quantum systems and further applications of such systems in quantum information processing as well as quantum communication.

\end{document}